\documentclass[a4paper,11pt]{article}
\pdfoutput=1
\usepackage{jheppub}
\usepackage[T1]{fontenc}
\usepackage{amsfonts}
\usepackage{amsmath}
\usepackage{amssymb}
\usepackage{multirow}
\usepackage{graphicx}
\usepackage{mathtools}
\usepackage{comment}
\usepackage{subcaption}

\def\be{\begin{equation}}
\def\ee{\end{equation}}
\def\bea{\begin{eqnarray}}
\def\eea{\end{eqnarray}}

\newcommand{\tr}{{\rm {tr}}}

\newcommand{\Tr}{{\rm {Tr}}}

\numberwithin{equation}{section}

\begin{document}

\title{Finite-$N$ Bootstrap Constraints in Matrix and Tensor Models}

\author{Samuel Laliberte}
\emailAdd{samuel.laliberte@oist.jp}
\author{and Reiko Toriumi}
\emailAdd{reiko.toriumi@oist.jp}
\affiliation{Okinawa Institute of Science and Technology Graduate University, 1919-1, Tancha, Onna, Kunigami District, Okinawa 904-0495, Japan.}

\date{\today}

\abstract{We explore how matrix bootstrap techniques can be used to constrain matrix and tensor models at finite $N$, where $N$ is the dimension of the matrix/tensor, taking a Gaussian model with a quartic interaction as example. For matrix models, we find further evidence that bounds do not depend explicitly on $N$, but rather on properties of multi-trace expectation values. For tensor models, the structure of the Schwinger-Dyson equations allow for bounds that vary as a function of $N$, admitting a broader scan of the parameter space of the theory. In the latter case, we find novel bounds on the two-point function as a function of the quartic coupling of the theory.}

\maketitle

\section{Introduction}

Random matrix models have a wide range of applications to diverse topics \cite{BoseHuZhou2019}, such as nuclear physics \cite{Wigner1955CharacteristicVO}, integrable systems \cite{Harnad2011RandomMatrices}, string theory \cite{Brezin:1990rb}, quantum chromodynamics \cite{tHooft:1973alw}, quantum chaos \cite{Brzin1996SpectralFF}, free probability \cite{Speicher2014FreePA, Lehner2018JAMR}, and data analysis \cite{Speicher2023HDARML}. Of particular notable result of random matrix models is that of 2-dimensional quantum gravity \cite{Brezin:1977sv, DiFrancesco:1993cyw, Douglas:1989ve}. Discrete representation of random planar maps was developed by random matrix models, and such  random planar maps made out of any polygons of the same size converge in distribution to the universal continuous random metric space called the Brownian sphere \cite{Gall2011UniquenessAU}. Moreover, it has been rigorously established \cite{miller2019liouvillequantumgravitybrownian, Miller2016LiouvilleQGII, Miller2016LiouvilleQGIII} that the Brownian sphere is equivalent to Liouville quantum gravity \cite{POLYAKOV1981207, David2014LiouvilleQG}, which provides the continuum description of two-dimensional quantum gravity coupled to conformal matter \cite{David1988CONFORMALFT, Distler:1988jt}.

In random geometric approach to quantum gravity, schematically one expects to have a partition function of the form
\be
Z \sim \sum_{\text{topologies}} 
\int {\mathcal{D}}g \, {\mathcal{D}}X \, e^{-S}\,,
\ee
where a path integral is taken over some geometric degrees of freedom including topologies and metric $g$ in addition to matter fields $X$, with appropriate weights $S$. In general, the above expression may be ill-defined, and to regulate it, one solution is to sum over discrete geometries. However, eventually, one needs to address going back to the continuum. In two dimensions, a theory of random matrix models provides such a successful description of the approach explained above. Random matrix models provide a generating function of ribbon graphs which are dual to topological piecewise-linear (PL) surfaces. Since there is still a leftover degree of freedom to assign metric to PL topological surfaces, one can for example assign the same length to each $1$-cell, and call it a lattice spacing, $a$, i.e., discrete surfaces are taken to be equilateral. One can organize the partition function by expanding it in small parameter $1/N$ in the large $N$ limit. Such is a topological expansion indexed by genus, which is a universal feature of random matrix models; at leading order in $1/N$ expansion, planar graphs dominate.  Tuning the coupling constants of the random matrix models to criticality (and sending $a \rightarrow 0$ such that the total area of the 2-surface is finite) yields infinite graphs with fixed genus at any order in $1/N$, therefore the theory is taken to be continuum at criticality. If instead, double scaling limit is taken, where $N \rightarrow \infty$ and tuning coupling constant to criticality is taken simultaneously in a correlated manner, not only the sphere but all the topologies contribute at the continuum limit. It is noteworthy to comment that including the equilateral choice, the simplest choices in general play sufficient and universal features are drawn for the description of $2$-dimensional quantum gravity provided  by random matrix models.

In higher dimensions, there have been similar attempts to describe quantum gravity via random geometric approach using random tensor models. Random tensor models are probability theory of random tensors of size $N^D$ and provide generating functions of $(D+1)$-edge-colored graphs which are proven to be dual to $D$-dimensional simplicial PL pseudo-manifolds \cite{Gurau:2013gya, Ferri1986AGR,Bandieri1982Generating}. In general, one would like to obtain a theory of random geometry in the limit $N \rightarrow \infty$, where the theory is analytically tractable. In this limit, similarly to the matrix case, $1/N$ reorganization of the partition function is governed by a topological and triangulation dependent quantity called Gurau degree, which reduces to the genus if $D=2$. At the leading order in $1/N$ expansion, graphs called melons \cite{Gurau:2011kk}, a particular triangulation subclass of the $D$-dimensional sphere, dominate. Melonic graphs are particularly nice as they are exactly analytically solvable. However, they behave as branched polymers (tree-like) in the continuum limit \cite{Bonzom:2011zz, Gurau:2013cbh}, therefore presently pathological for quantum gravity. Beyond the melonic limit, analytic options are often limited to solve the theory. Given that understanding the behavior beyond the melonic sector is significantly more challenging, and analytic control appears limited, numerical investigations may be required to explore the regimes relevant for higher dimensional gravity.

Lately, the bootstrap approach has been useful to study matrix models numerically. This approach was first developed for matrix integrals \cite{Lin:2020mme,Kazakov:2021lel,Khalkhali:2020jzr} were it proved useful to probe critical behavior of the theory. It was then extended to matrix quantum mechanics \cite{Han:2020bkb}, where it proved useful to study the low energy behavior of theories. An important ramification of this program has been the study of matrix models related to string theory (see \cite{Lin:2025iir} for a review). There, the bootstrap approach has led to novel constraints on the BFSS matrix model \cite{Lin:2023owt,Lin:2024vvg,Lin:2025srf}, which is closely tied to the dynamics of D$_0$ branes in string theory. The IKKT model \cite{Li:2025tub} and the Marinari-Parisi model \cite{Laliberte:2025xvk} have also received attention. Other notable excursions of the matrix bootstrap include the study of matrix quantum mechanics at finite temperature \cite{Cho:2024kxn,Cho:2025nlv,Adams:2025nww}, Euclidean two-point functions of matrix models \cite{Cho:2025vws}, matrix models and their eigenvalue distribution \cite{Kovacik:2025qgj}, matrix models with a complex probability measure \cite{Maeta:2026oku}, and unitary matrix integrals \cite{Berenstein:2026wky}.

Despite recent advances in the matrix bootstrap program, some research directions remain largely unexplored. This is the case, for example, for matrix models at finite $N$. For the bootstrap approach to yield strong bounds, one must usually take the large $N$ limit and make use of large $N$ factorization. At finite $N$, factorization breaks down and one must consider new tools. Recent work has shown some progress in overcoming this problem. Notably, in \cite{Kazakov:2024ool}, it was shown that multi-trace relations specific to certain values of $N$ can lead to bounds on simple matrix models transforming under the SU($N$) group, with the example of SU(2) and SU(3) worked out explicity\footnote{ This topic has also been studied for matrix quantum mechanical models transforming under the SU($N$) group \cite{Cho:2024kxn}. In this case, bounds appear naturally at finite $N$ without using multi-trace relations.}. However, finite $N$ constraints on models obeying other symmetry groups remain an open topic of study. Another direction that remains to be explored is applications to tensor models. Matrix models are a sub-family of the broader family of tensor models. Therefore, it is interesting to consider whether matrix bootstrap techniques also apply to this wider variety of models. Moreover, the connection between tensor models and random surfaces in higher dimensions is still an active topic of study. Understanding how bootstrap methods apply to tensor models may prove helpful to explore this topic.

In the present work, we explore how matrix bootstrap techniques can be used to constrain matrix and tensor models at finite $N$, where $N$ is the dimension of the matrix/tensor models. Our paper is structured as follows. In Section \ref{sec:matrix_positivity_constraints}, we write down the matrix bootstrap techniques that will be used in our anaylsis. In Section \ref{sec:matrix_boot}, we test these techniques on a Gaussian matrix model with a quartic interaction.  Finally, in Section \ref{sec:tensor_boot}, we show how matrix bootstrap techniques allow for bounds on a Gaussian tensor model with a quartic interaction.

\section{Positivity constraints}
\label{sec:matrix_positivity_constraints}

Before showing exemplary uses of the bootstrap method for matrix and tensor models, let us first write down the positivity constraints that we will use throughout the present work. Each of these constraints follow a simple reasoning, which is that the expectation value of a strictly positive quantity computed using a probability distribution with positive norm must be a strictly positive number. By imposing this reasoning on various matrix-valued quantities, one can obtain various positivity constraints that can be used to find bounds on these matrix-valued quantities. We enumerate the relevant examples below.

\vspace{1em}

\noindent
\textbf{Single-trace positivity:} The first constraint we will discuss is {\it single-trace positivity}, which was first introduced in \cite{Lin:2020mme}. The idea behind this constraint is to start by considering an arbitrary complex matrix $\Phi$. Assuming a positive probability measure\footnote{Throughout the present paper, we will assume that we are dealing with models that have a positive probability measure. In the present case, this means that the potential of the matrix or tensor model must be a real quantity. Some models with a complex probability measure may violate the present assumptions, and require a different type of analysis (see \cite{Maeta:2026oku,Berenstein:2026wky} for recent developments). We leave the study of these models to future work.}, the statement for the expectation value of a single-trace invariant
\be
\label{eq:singletrpos}
\langle \tr \Phi^\dagger \Phi \rangle \geq 0 \, ,
\ee 
must be true. This is just the statement that a strictly positive number averaged over a positive probability measure must give a positive quantity. Here, we will use the notation that the weighted trace $\tr \equiv \frac{1}{N} \Tr$ is the usual trace $\Tr$ weighted by a factor of $N$. This will be done so that the weighted trace of the identity matrix $\mathbf{1}$ yields $\tr \mathbf{1} = \frac{1}{N} \Tr \mathbf{1} = 1$. One could then imagine letting this matrix be a linear combination 
\be
\Phi = \sum_i c_i \mathcal{O}_i
\ee
of arbitrary complex matrices $ \mathcal{O}_i$ weighted by arbitrary complex coefficients $c_i$. The inequality \eqref{eq:singletrpos} then reads
\be
\langle \tr \Phi^\dagger \Phi \rangle = \sum_{i, \, j} c_i^\dagger \langle \tr \mathcal{O}_i^\dagger \mathcal{O}_j \rangle c_j \geq 0 \, .
\ee
If the inequality above must be true for any choice of coefficients $c_i$, we obtain that the Gram matrix
\be
\mathcal{M}_{ij} = \langle \tr \mathcal{O}_i^\dagger \mathcal{O}_j \rangle \, ,
\ee 
must be positive semi-definite ($\mathcal{M} \succeq 0$). In a matrix bootstrap context, single-trace positivity constraint can be used to constrain the possible values of a set of single-trace observables. For example, for a Hermitian one-matrix model, one is usually interested in the set of powers $\mathcal{O}_k = M^k$ of a matrix $M$. In this case, the Gram matrix takes the form
\be
\mathcal{M}_{ij} = \langle \tr M^{i+j} \rangle \, .
\ee
For multi-matrix models, one usually constrains the set of all possible "words" that can be made from a combination of the matrices. For example, for a two matrix model with the matrices $A$ and $B$, one would consider the matrix vector $\mathcal{O}_i = (1 , A , B, A^2, AB, BA, B^2, ...)$ containing all the possible "words" that can be made from the letters $A$ and $B$ up to a certain word length $l$.

\vspace{1em}

\noindent
\textbf{Double-trace positivity:} A statement similar to the one above, also introduced in \cite{Lin:2020mme}, can be used to constrain double-trace observables. Let us consider a complex scalar $P$. For this complex scalar, the statement
\be
\langle P^\dagger P \rangle \geq 0 \, 
\ee
is known to be true for a positive probability measure. This is just saying that the expectation value of the norm of a complex variable must be positive. Now, let us assume that $P$ is constructed by a sum of single-trace variables of the form
\be
P = \sum_i c_i \tr \mathcal{O}_i \, .
\ee
Here, $c_i$ are complex coefficients, and the $\mathcal{O}_i$'s are complex matrices. In this case, the statement that  the expectation value of the norm of $P$ must be positive reads
\be
\langle P^\dagger P \rangle = \sum_{ij} c_i^\dagger \langle \tr \mathcal{O}_i^\dagger  \tr \mathcal{O}_j \rangle c_j \geq 0 \, .
\ee
If this is true for any values of $c_i$, we arrive at the statement that the Gram matrix
\be
\mathcal{Q}_{ij} =  \langle \tr \mathcal{O}_i^\dagger \tr \mathcal{O}_j \rangle \, ,
\ee
must be positive semi-definite ($\mathcal{Q} \succeq 0$). For a hermitian one-matrix model where we choose $\mathcal{O}_k = M^k$, the Gram matrix above would take the form
\be
\mathcal{Q}_{ij} =  \left\langle \tr M^i \tr M^j \right\rangle \, ,
\ee
hence constraining products of single-trace quantity $\tr M^k$. For two-matrix models, {\it double-trace positivity}, as we will name it, constrains products of a trace $\tr \mathcal{O}_i$ of words encoded in a matrix vector $\mathcal{O}_i = (1 , A , B, A^2, AB, BA, B^2, ...)$.

\vspace{1em}

\noindent
\textbf{Traceless component positivity:} We will finally introduce a new constraint named {\it{traceless component positivity}}. It is known that any matrix $M$ can be decomposed as $M = M_{\text{traceless}} + \tr M \, \mathbf{1}$ where $M_{\text{traceless}} = M - \tr M \, \mathbf{1}$ is the traceless part of the matrix, and where $\mathbf{1}$ is the identity matrix. Using this fact, it is useful to consider how the present positivity statements can be defined for the traceless part of matrix observables. Let us define the traceless complex matrix $(\mathcal{O}_\text{traceless})_i \equiv \mathcal{O}_i - \tr \mathcal{O}_i 
\, \mathbf{1}$ associated to a vector of operators $\mathcal{O}_i$. Imposing that single-trace positivity holds for $(\mathcal{O}_\text{traceless})_i$ implies that the condition
\be
\langle \tr (\mathcal{O}^\dagger_\text{traceless})_i (\mathcal{O}_\text{traceless})_j \rangle \succeq 0 \, ,
\ee 
must be satisfied. Expanding this expression, we find that this is equivalent to imposing that the condition $\mathcal{M} - \mathcal{Q} \succeq 0$ must be satisfied for $\mathcal{O}_i$. Although it follows directly from single-trace positivity, we found that imposing the constraint above led to stronger constraints when used conjointly with $\mathcal{M} \succeq 0$ and $\mathcal{Q} \succeq 0$. This is also what was observed in \cite{Lin:2025srf}, where a similar statement was made to constrain the quantity $\mathcal{O}_i - \left\langle \tr \mathcal{O}_i \right\rangle \mathbf{1}$. 
In this case, it was also found that imposing $\big\langle \tr \left( \mathcal{O}_i - \left\langle \tr \mathcal{O}_i 
\right\rangle \mathbf{1} \right)\left(\mathcal{O}_j - \left\langle \tr \mathcal{O}_j \right\rangle \mathbf{1}
\right)\big\rangle 
\succeq 0$ leads to stronger constraints. The main difference between the present statement and the condition explored in \cite{Lin:2025srf} is that traceless component positivity involves double-trace expectation values. This will be useful when exploring bounds at finite $N$, where double-trace expectation values do not factorize.

\section{Bounds on matrix models}
\label{sec:matrix_boot}

Using the positivity statements of the previous section, we will now show how to constrain matrix models. As an example case, let us study the Gaussian matrix model with quartic interaction described by the following partition function
\be
\label{eq:quarticmmpartitionfcn}
Z = \int dM e^{-N \, \Tr V(M)} \quad , \quad V(M) = \frac{1}{2} M^2 + \frac{g}{4} M^4 \, .
\ee
Here, $M$ is taken to be a Hermitian matrix. To constrain the system above, we will impose positivity constraints along with the Schwinger-Dyson equations for the system. The Schwinger-Dyson equations can be found by invoking that the expectation value
\be
\langle \mathcal{O}_{ab}(M) \rangle = \frac{1}{Z} \int dM \, \mathcal{O}_{ab}(M) \, e^{- N \, \Tr V(M)}
\ee 
of an operator ${\mathcal{O}}_{ab}(M)$ depending on $M$ (in index notation with $a, b = 1, 2, ...,N$) is invariant under a infinitesimal change of variables $M \rightarrow M + \delta M$. This leads us to the conditions
\be
\frac{1}{Z} \sum_{a, \, b} \int dM  \frac{\partial}{\partial M_{ab}} \left[ \mathcal{O}_{ab}(M) \, e^{- N \, \Tr V(M)} \right] = 0 \, ,
\ee
which gives rise to the following set of equations
\be
\sum_{a, \, b} \left\langle \frac{\partial \mathcal{O}_{ab}(M)}{\partial M_{ab}} \right\rangle =  N 
\sum_{a, \, b} 
\left\langle \mathcal{O}_{ab}(M) \frac{\partial V(M)}{\partial M_{ab}} \right\rangle \, . 
\label{eq:matrix_sde}
\ee
In the present case, we will be interested in the Schwinger-Dyson equations that arise from the family of operators $\mathcal{O}$ that can be expressed as a power $\mathcal{O} = M^l$ of the matrix $M$. In this case, \eqref{eq:matrix_sde} reduces to
\be
\label{eq:mmsde}
\sum_{k=0}^{l-1} \langle \tr M^{k} \, \tr M^{l-k-1} \rangle = \langle \tr M^{l+1} \rangle + g \langle \tr M^{l+3} \rangle 
\ee
Here, we have divided the entire expression by a factor of $1/N^2$ so that each term can be written using the weighted trace $\tr$.
 
\subsection*{Bounds at large $N$}
\label{sec:largeNboundsMM}

\begin{figure}[h]
	\centering
	\begin{minipage}{0.45\textwidth}
		\centering
		\includegraphics[width=\linewidth]{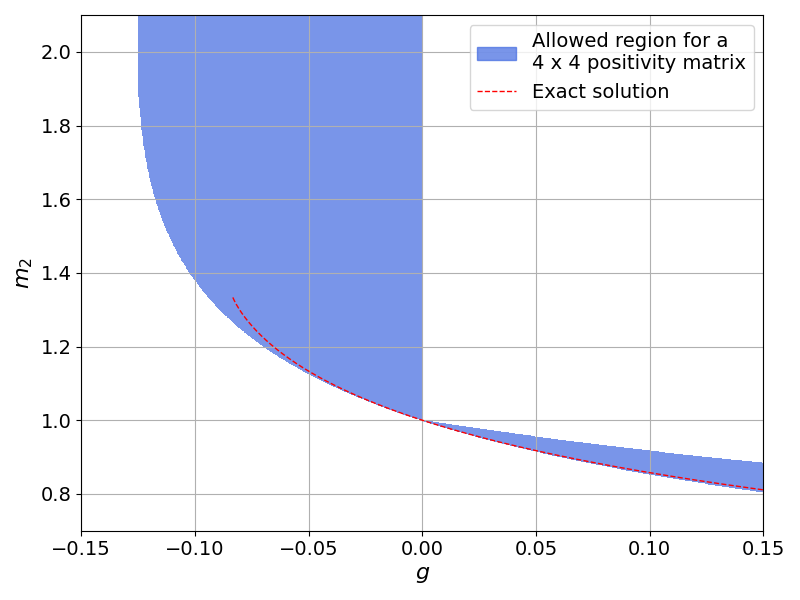}
	\end{minipage}\hfill
	\begin{minipage}{0.45\textwidth}
		\centering
		\includegraphics[width=\linewidth]{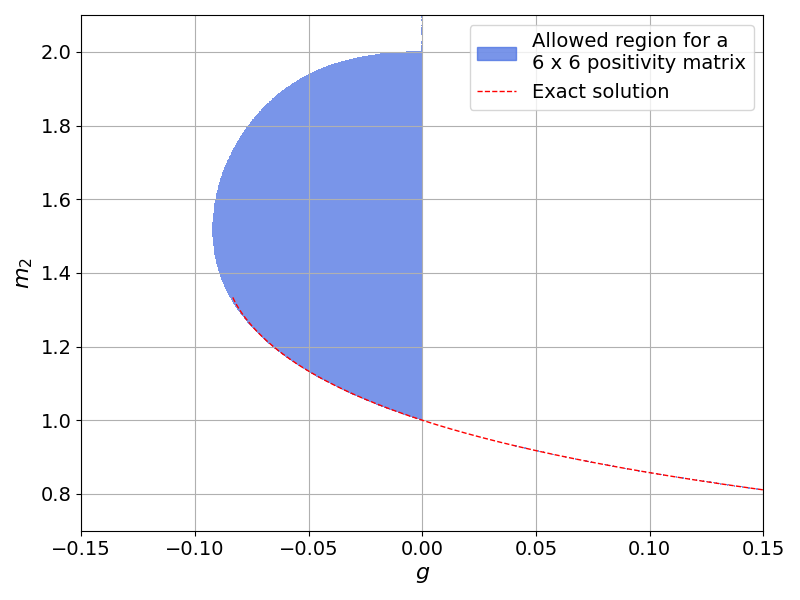}
	\end{minipage}\hfill
	\begin{minipage}{0.45\textwidth}
		\centering
		\includegraphics[width=\linewidth]{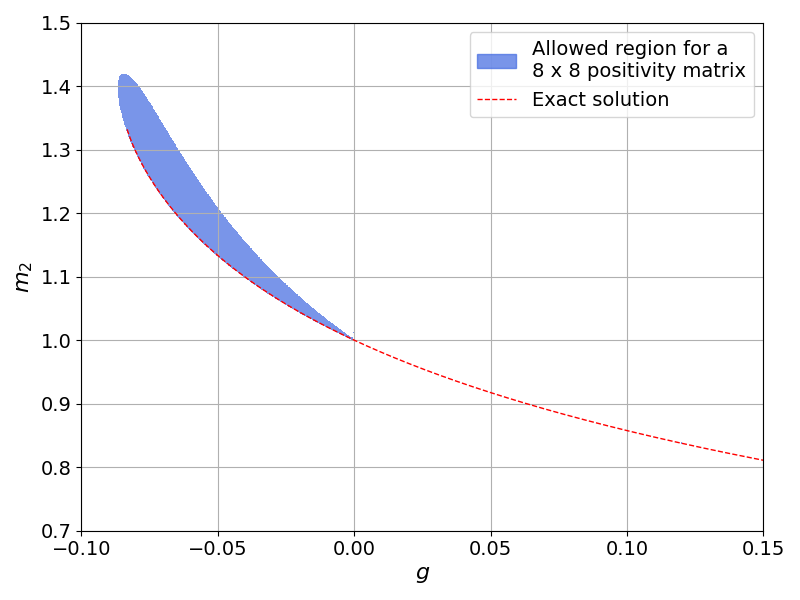}
	\end{minipage}\hfill
    \begin{minipage}{0.45\textwidth}
		\centering
		\includegraphics[width=\linewidth]{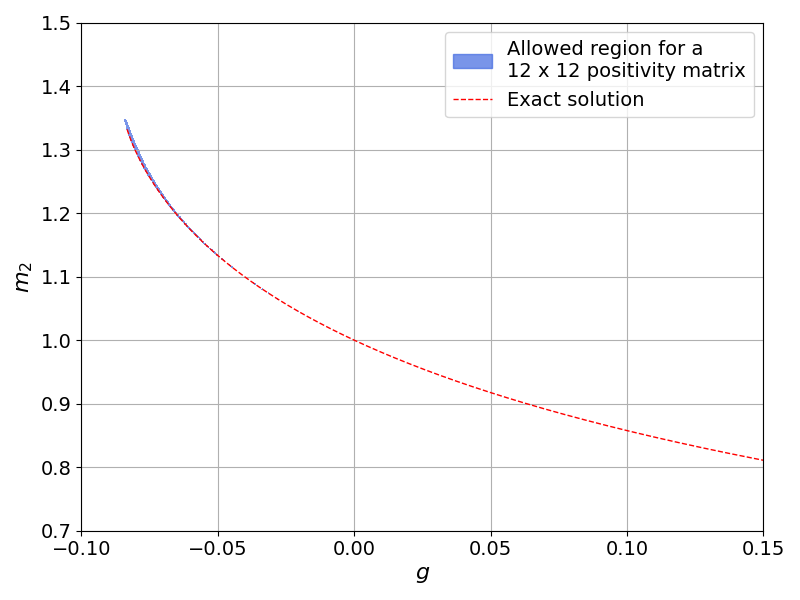}
	\end{minipage}\hfill
	\caption{Bounds for the two-point function $m_2$ as a function of the coupling $g$. The size of the Gram matrix is varied from $4 \times 4$ (top left), to $6 \times 6$ (top right), $8 \times 8$ (bottom left), and finally $12 \times 12$ (bottom right).}
	\label{fig:M4_largeN_bounds}
\end{figure}

Let us first demonstrate how the bootstrap approach works by considering the large $N$ limit of the theory. At large $N$, the factorization property
\be
\lim_{N \rightarrow \infty} \langle \tr M^k \tr M^l \rangle = \langle \tr M^k \rangle \langle \tr M^l \rangle \, 
\label{eq:largeNfact}
\ee
applies. Defining the quantity $m_k \equiv \langle \tr M^k \rangle$, the Schwinger-Dyson equations \eqref{eq:mmsde} can be expressed as a function of single-trace expectation values as follows
\be
\sum_{k=0}^{l-1} m_{k} m_{l-k-1} = m_{l+1} + g \, m_{l+3} \, .
\label{eq:M4_sde_largeN}
\ee
In the present limit, the Schwinger-Dyson equations \eqref{eq:largeNfact} can be solved exactly using the generating function method (See \cite{Kazakov:2021lel} for an example use of this method). For the two-point function of the system, we obtain
\be
m_2 = \frac{4 a - a^2}{3} \, .
\ee
where $a = - \frac{1}{6 g} \left(1 - \sqrt{1+12 g}\right)$ \cite{Khalkhali:2020jzr}. It is possible to check that the bootstrap approach allows us to recover bounds consistent with the result above.  Let us consider the bounds on the two-point function as an example. In the present case, the recursion relation defined by the set of Schwinger-Dyson equations in the large $N$ limit 
\eqref{eq:largeNfact}
depends only on two variables: $m_2$ and $g$. Using these equations and single-trace positivity, it is possible to obtain bounds on $m_2$ as a function $g$. To do so, we use the Gram matrix
\be
\mathcal{M}_{ij} = m_{i + j} \, ,
\ee
and substitute in \eqref{eq:M4_sde_largeN} recursively for even moments, setting the odd moments to zero as imposed by the symmetry of the system. Since $\mathcal{M}$ only depends on $m_2$ and $g$, we can directly test if $\mathcal{M}$ is positive semi-definite using a 2D grid of $m_2$ and $g$ values in a certain interval to obtain bounds on the system. Increasing the size of the Gram matrix $\mathcal{M}$ then allows us to recover progressively stronger bounds on $m_2$ vs $g$ (see Figure \ref{fig:M4_largeN_bounds}).

As we can see on Figure \ref{fig:M4_largeN_bounds}, the bounds (blue region) start to get very close to the exact solution (dotted red line) for a Gram matrix of size $8 \times 8$. When increasing the size of the Gram matrix beyond $8 \times 8$, the bounds close in even tighter on the exact solution, excluding any other possible values of $m_2$ and $g$ on the $m_2$-$g$ plane. Therefore, the bootstrap bounds provide a good approximation of the exact solution if the Gram matrix is sufficiently large.

\subsection*{Bounds at finite $N$}

Let us now use the positivity bounds of Section \ref{sec:largeNboundsMM} to extend the bootstrap method to the case of finite $N$. In this case, large $N$ factorization no longer holds and we must add a new variable $m_{k,l} \equiv \langle \tr M^k \tr M^l \rangle$ related to double-trace expectation values. The Schwinger-Dyson equations 
\eqref{eq:mmsde}
then read
\be
\sum_{k=0}^{l-1} m_{k,l-k-1} = m_{l+1} + g \, m_{l+3} \, .
\label{eq:M4_sde_finteN}
\ee
Because the Schwinger-Dyson equations no longer only depend on $m_2$ and $g$, but also on new variables $m_{k,l}$ related to the double traces, the "guess-and-check" method of 
Section \ref{sec:largeNboundsMM}
quickly becomes computationally expensive as one increases the size of the Gram matrix. For this reason, it is better to treat the present system as a convex optimization problem to find the extrema of the allowed region for $m_2$ at a fixed value of $g$, and infer the allowed region from these extrema. The idea, which was developed in \cite{Kazakov:2021lel}, is as follows. Let $v_i = (m_2 , m_{1,1}, m_{1,2}, ...)$ be a vector containing all unknown moments in the theory. The present problem consists in finding the extremum of the loss function $f = m_2$ such that the Schwinger-Dyson equations \eqref{eq:M4_sde_finteN} and the positivity constraints are satisfied. This problem can be solved either using Mathematica or other available semi-definite programming codes. In the present paper, we used SDPA \cite{Yamashita2012SDPA} since it provides a good balance between accuracy and execution speed. Moreover, we used a coupling rescaling trick to improve the accuracy of our results, especially in the small $g$ regime (see Appendix \ref{sec:coupling_rescaling}). 

\begin{figure}[h]
	\centering
	\begin{minipage}{0.5\textwidth}
		\centering
		\includegraphics[width=\linewidth]{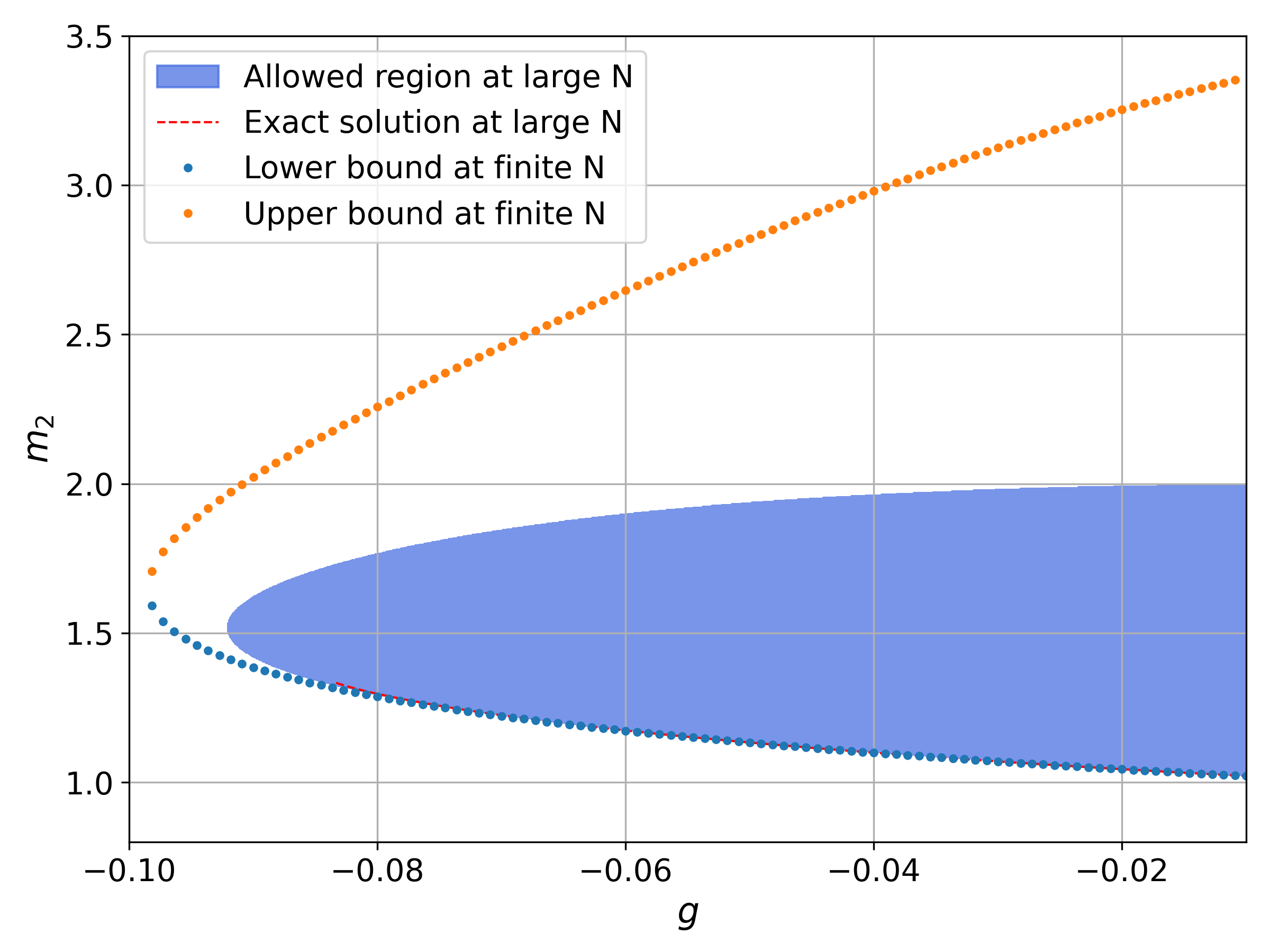}
	\end{minipage}\hfill
	\begin{minipage}{0.5\textwidth}
		\centering
		\includegraphics[width=\linewidth]{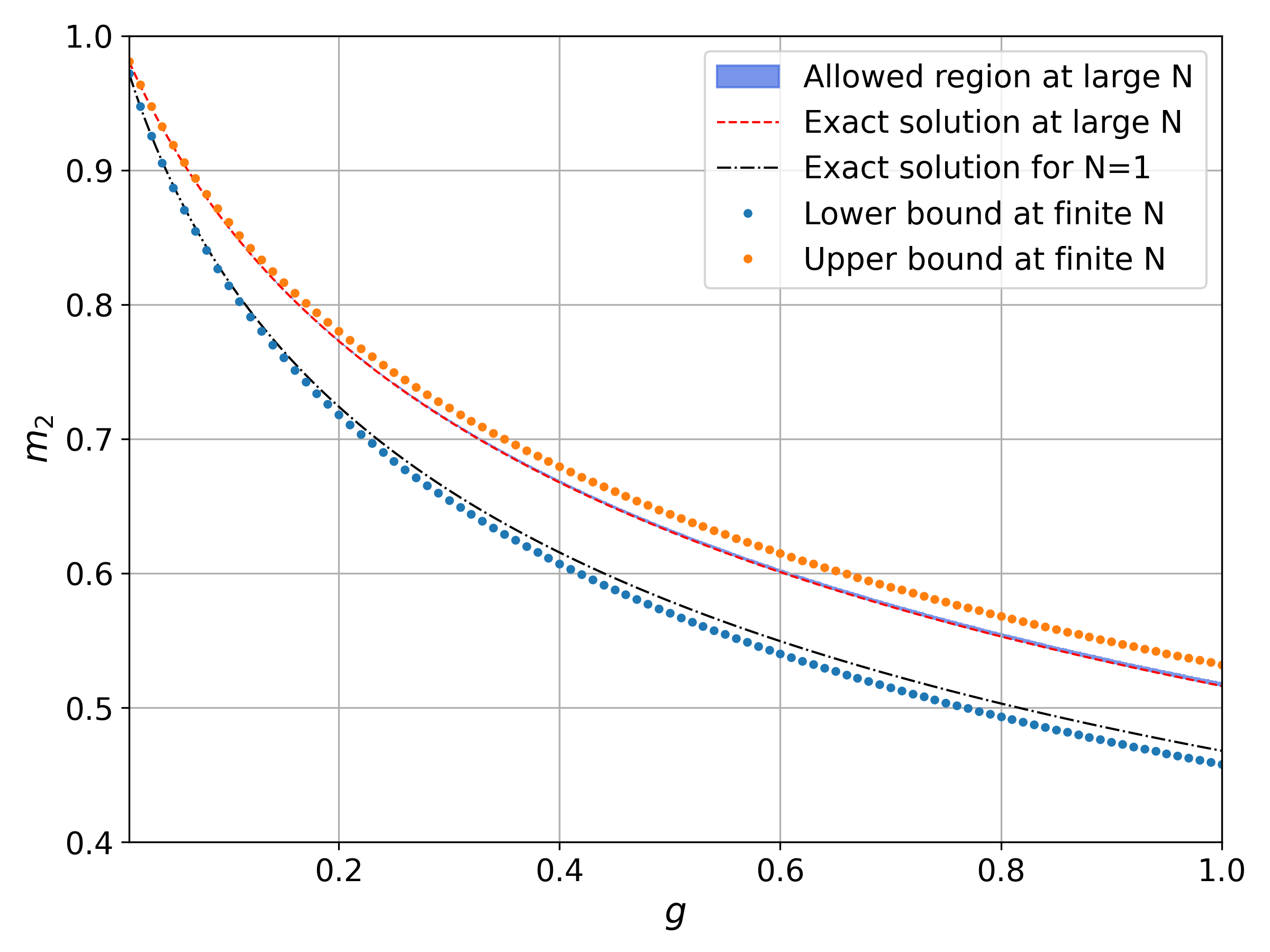}
	\end{minipage}\hfill
	\caption{Bounds on $m_2$ vs $g$ at finite $N$ for a $6 \times 6$ Gram matrix. The figure on the left shows bounds for $g < 0$ and the figure on the right shows the bounds for $g > 0$.}
	\label{fig:M4_finiteN}
\end{figure}

By solving this semi-definite programming problem, we obtained the results of Figure \ref{fig:M4_finiteN}. In this figure, we found that the bounds at finite $N$ are weaker than the bounds at large $N$ since we do not have large $N$ factorization, and are consistent with both the $N = 1$ and the large $N$ limit. This seems to follow directly from the fact that neither the Schwinger-Dyson equations nor the positivity conditions depend directly on $N$. Our interpretation is that in absence of explicit $N$ dependence, the bounds must account for "all" possible curves at finite $N$. This results in bounds that are consistent with both the $N = 1$ and the large $N$ limit.

Motivated by this apparent lack of explicit $N$ dependence, we tried restoring the explicit dependence of $N$ in the present problem by choosing the optimization variables $m_k = \frac{1}{N} \langle \Tr M^k \rangle$ instead of $m_k = \langle \tr M^k \rangle$. However, in this case, we did not find that the bounds for $m_k = \frac{1}{N} \langle \Tr M^k \rangle$ vs g varied as a function of $N$.

\begin{figure}[h]
	\centering
	\begin{minipage}{0.5\textwidth}
		\centering
		\includegraphics[width=\linewidth]{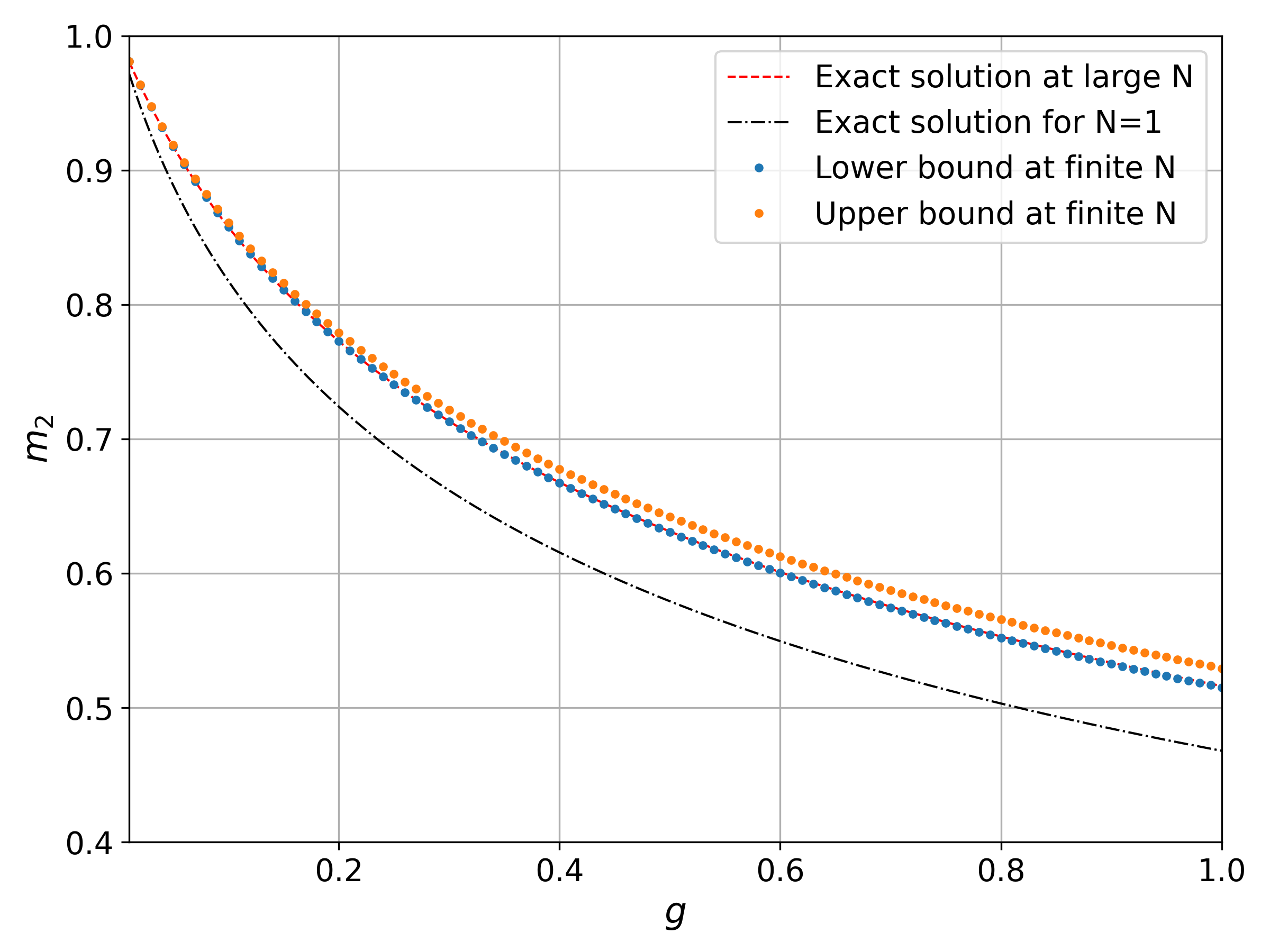}
	\end{minipage}\hfill
	\begin{minipage}{0.5\textwidth}
		\centering
		\includegraphics[width=\linewidth]{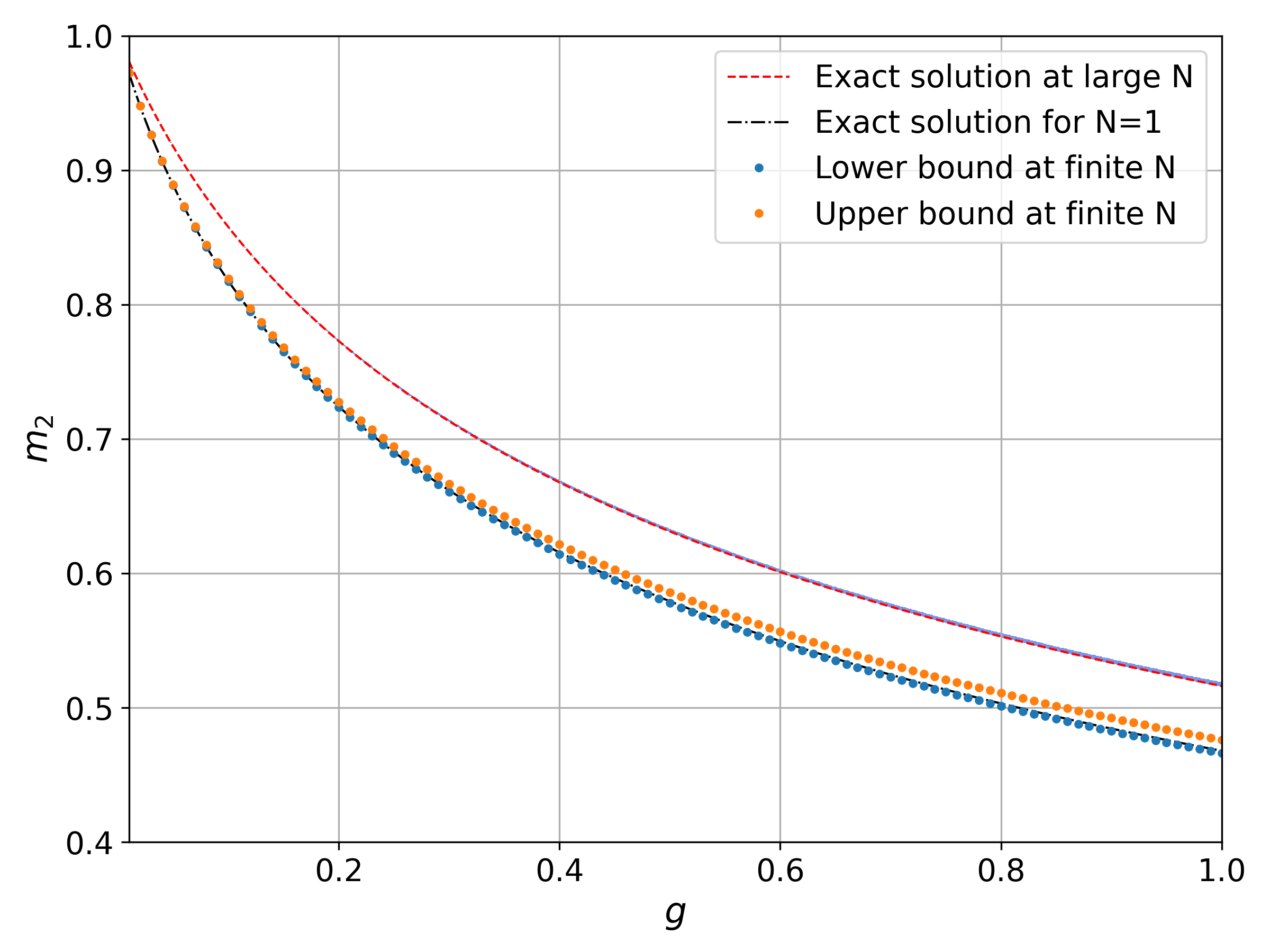}
	\end{minipage}\hfill
	\caption{Bounds on $m_2$ vs $g$ at finite $N$ for a $6 \times 6$ Gram matrix. The figure on the left shows bounds when imposing the large $N$ factorization properties and the figure on the right shows bounds when imposing the $N = 1$ factorization properties.}
	\label{fig:M4_finiteN_largeNvsN=1}
\end{figure}

We then tried imposing various factorization properties for the double-trace expectation values and then found a good match with the $N = 1$ and large $N$ limit (see Figure \ref{fig:M4_finiteN_largeNvsN=1}). To recover the large $N$ limit, we imposed that $m_{k,l}$ must be zero if $k$ or $l$ is odd. This property arises naturally in the large $N$ limit as a consequence of large $N$ factorization and the parity constraints of the system. When imposing this additional constraint, the bounds close in on the exact solution in the large $N$ limit, excluding the exact solution for $N = 1$. To recover the $N = 1$ limit, we then imposed that $m_{k,l} = m_{k + l}$, which follows directly from the product property $x^k x^l = x^{k+l}$ recovered in the $N = 1$ limit. In this case, we found that the bounds close in on the exact solution for $N = 1$, excluding the exact solution in the large $N$.  

The fact that the bounds do not depend explicitly on $N$, while the $N=1$ and large $N$ limits can be recovered by imposing properties of multi-trace expectation values, suggests that the $N$ dependence of the present system is encoded in specific properties of multi-trace expectation values. This is consistent with the expectation that bootstrap bounds at finite $N$ depend on such properties \cite{Kazakov:2024ool}. One might therefore expect that additional constraints on multi-trace expectation values, similar to those explored in \cite{Kazakov:2024ool}, could lead to stronger bounds at particular values of $N$. It would be interesting to explore this perspective in future work.

In the present analysis, we found that the bounds did not improve in strength as the size of the $n \times n$ Gram matrix was taken beyond $n = 6$. Our hypothesis for why this is the case is that beyond $n = 6$, unconstrained values of $G_{n,m}$ prevent improvements of the constraints. For $n \leq 6$, all values of $G_{n,m}$ that appear in $\mathcal{M}$ also appear in $\mathcal{Q}$, and are therefore constrained by the inequalities $\mathcal{M} \succeq 0$, $\mathcal{Q} \succeq 0$, and  $\mathcal{M} - \mathcal{Q} \succeq 0$. However, for $n \geq 6$, some values of $G_{n,m}$ that appear in $\mathcal{M}$ do not appear in $\mathcal{Q}$. For this reason, they become only constrained by the inequalities $\mathcal{M} \succeq 0$ and $\mathcal{M} - \mathcal{Q} \succeq 0$, giving rise to weaker constraints. Additional constraints may be required to get stronger results past $n = 6$.

\section{Bounds on tensor models}
\label{sec:tensor_boot}

We will now show how the formalism of Section \ref{sec:matrix_boot} applies to tensor models. Before digging into this topic, we will review some basics of tensor models, highlighting key distinctions with the matrix case. We will then apply the bootstrap methods to a tensor model with a quartic interaction.

In the study of tensor models, one is usually interested in the free energy $F$ related to the partition function
\be
Z = e^{- N^D F}= \int d T d \bar T e^{- N^{D-1} 
\,
S(T,\bar T)} \, ,
\ee
which relates to a tensor model with a potential of the form 
\be
S(T , \bar T) = T \cdot \bar T + \sum_i t_i B_i(T , \bar T) \, .
\label{eq:tensor_pot}
\ee
Here, $T \cdot \bar T = \sum_{\vec{a}} T_{\vec{a}} \bar T_{\vec{a}}$ is short notation for a product of two tensors $T = T_{\vec{a}}$ and $\bar T = \bar T_{\vec{a}}$ where the indices contained in the $D$-dimensional index vector $\vec{a} = (a_1 , a_2 , ... , a_D)$ are fully contracted with each other.

Like matrix models, where the potential contains single-trace observables invariant under unitary transformations $M \rightarrow U^\dagger M U$, tensor models contain "bubbles" $B_i$, which are invariant under unitary transformations of the form
\be
T_{a_1  a_2 ...  a_D} \rightarrow \sum_{b_1 , b_2 , ... , b_D} U^{(1)}_{a_1  b_1} U^{(2)}_{a_2  b_2} ... U^{(D)}_{a_D  b_D} T_{b_1  b_2  ...  b_D} \,,
\ee 
where $U^{(i)}$ are independent unitary matrices of size $N \times N$. The simplest example of a bubble is the elementary bubble $T \cdot \bar T$ that arises at quadratic order in the tensor potential \eqref{eq:tensor_pot}. This elementary bubble is often referred to as the {\it elementary melon} in the tensor model literature. The other bubbles can be found by fully contracting the indices of a tensor monomial $T \bar T T ...$ in a way that respects the order of the indices ($a_1$ contracted with $b_1$ and so forth).

From the free energy, or directly from the probability distribution of the tensor model, one usually computes the expectation value of these bubbles, defined as
\be
\langle B_i(T , \bar T) \rangle = \frac{1}{Z} \int d T d \bar T 
\,
B_i(T , \bar T)
\,
e^{- N^{D-1} S(T , \bar T)} \, ,
\ee
using perturbative methods. In the present case, we will constrain the possible values of these expectation values using bootstrap methods, and compare with exact results.

For our analysis, we will restrict ourselves to a subclass of tensor models that can be mapped onto a rectangular matrix model \cite{Gurau:2011sk}. In these models, one introduces a "fat" index $A = (a_2 , a_3 , ... , a_D)$ of size $D-1$ which encodes all $N^{D-1}$ possible values of the vector made of the indices $a_2 , a_3 , ... , a_D$. Using this notation, the tensor $T_{a A}$ can be viewed as a $N \times N^{D-1}$ matrix where $a$ has $N$ elements and $A$ has $N^{D-1}$  elements. Similarly, its complex conjugate $\bar T_{a A} = (T^\dagger)_{A a}$ can be viewed as the $N^{D-1} \times N$ dimensional hermitian conjugate of $T$. For this class of model, the most general action one can write takes the form
\be
\label{eq:actioncyclic}
S(T , T^\dagger) = \Tr (T T^\dagger) + \sum_{p = 2} t_p \Tr (T T^\dagger)^p
\,.
\ee
In the above, $(T T^\dagger)_{ab}$ is the $N \times N$ matrix obtained by contracting the fat indices of $T$ and $\bar T$ together, leaving the lower case indices $a$ and $b$ open. Moreover, the trace $\Tr (T T^\dagger)$ is obtained by further contracting the lower-case indices $a$ and $b$ together.

More specifically, we will be interested in the simplest non-Gaussian model in the class of models above \eqref{eq:actioncyclic}.
The action of this model simply reads
\be
\label{eq:actionpillow}
S(T , T^\dagger) = \Tr (T T^\dagger) + g \, \Tr (T T^\dagger)^2 \, ,
\ee
where the last term plays the role of a quartic interaction\footnote{This specific interaction term is often referred to as a pillow interaction in the tensor model literature.} with a quartic coupling $t_2 = g$. Like for the matrix case, we can obtain bounds on this model using the positivity statements of Section \ref{sec:matrix_positivity_constraints}. To do so, we first impose the constraints arising from the Schwinger--Dyson equations. These follow from the invariance of the expectation value
\be
\langle \mathcal{O}_{aA}(T,T^\dagger)\rangle = \frac{1}{Z}\int dT \, dT^\dagger \, \mathcal{O}_{aA}(T,T^\dagger) \, e^{-N^{D-1}S(T,T^\dagger)}
\ee
under an infinitesimal change of variables \(T \rightarrow T + \delta T\). In the present case, we will be interested in the constraints related to the operator $\mathcal{O}_{aA}(T,T^\dagger) = \big[(TT^\dagger)^i T\big]_{aA}$. This leads us to the conditions
\be
\frac{1}{Z} \sum_{a A} \int dT d T^\dagger \frac{\partial}{\partial T_{a A}} \left( [(T T^\dagger)^i T]_{a A} \, e^{- N^{D-1} S(T , T^\dagger)} \right) = 0 \, .
\label{eq:sdeforrectangularmatrix}
\ee
The results of expanding the expression above, which where derived in \cite{Bonzom:2012cu}, can be written down as a function of two quantities, the first being the single-trace expectation value $G_k \equiv \big \langle \tr (T T^\dagger)^k \big \rangle$, the second being the double-trace expectation value $G_{k,l} \equiv \big \langle \tr (T T^\dagger)^k \tr (T T^\dagger)^l \big \rangle$. Making use of these two definitions, the Schwinger-Dyson equations \eqref{eq:sdeforrectangularmatrix} found from $\mathcal{O}_{a A}$  for the model \eqref{eq:actionpillow} simply reads
\be
G_l + \frac{1}{N^{D-2}} \sum_{k=0}^{l-1} G_{k,l-k} = G_{l+1} + 2 g \, G_{l+2}\,.
\label{eq:tensor_sdes}
\ee
Here, we have divided the Schwinger-Dyson equations by an overall factor of $1/N^{D}$ to use the weighted trace $\tr$ as defined in $G_k$ and $G_{k,l}$. 

The set of equations above \eqref{eq:tensor_sdes} is easily solvable in the large $N$ limit. In this limit, the $G_{k,l-k}$ term drops out because of the $1/N^{D-2}$ prefactor. Moreover, Gaussian universality \cite{Gurau:2011kk} implies $G_{2n} = G_1^n$. In this limit, the set of Schwinger-Dyson equations \eqref{eq:tensor_sdes} reduces to the simple expression
\be
1 = G_1 + 2 g G_1^2 \, .
\ee
Solving for $G_1$ in this expression, we obtain
\be
G_1 = \frac{-1 + \sqrt{1+8 g}}{4 g} \, . 
\label{eq:G1_largeN}
\ee
Another solvable limit for this model is the $N = 1$ limit. In this case, the value of $G_1$ can be simply integrated to obtain an expression depending on Bessel functions.

Glancing at the structure of the Schwinger-Dyson equations  \eqref{eq:tensor_sdes}, one can see similarities with the Schwinger-Dyson equations of the quartic matrix model \eqref{eq:M4_sde_finteN}. Namely, the equations involve single-trace expectation values and a sum over double-trace expectation values. Because of this structure, it is possible to define a bootstrap problem very similar to what we used to study matrix models at finite $N$. Consider, for example, the vector 
\be
\label{eq:wordsfortensor}
(O_i)_{ab} = (\delta_{ab} , M_{ab} ,  M^2_{ab}, ...) = M^i_{ab} \, ,
\ee
where the matrix $M_{ab}$ is defined as
\be
M_{ab} = (T T^\dagger)_{ab} \,, 
\ee
and the subscript $i$ on the l.h.s. of \eqref{eq:wordsfortensor} represent the component of the vector ${\mathcal{O}_i}$, but on the r.h.s. of \eqref{eq:wordsfortensor}, the superscript $i$ is the power of $M$. In particular, we have defined for $i=0$, $M_{ab}^0 \equiv \delta_{ab}$, where $\delta_{ab}$ is an identity matrix. In this case, imposing single-trace positivity of matrix models leads to the condition
\be
\label{eq:tensorGramcond}
\mathcal{M}_{ij} = \langle \tr O_i O_j \rangle = G_{i+j} \succeq 0 \, .
\ee 
Similarly, imposing double-trace positivity leads to the condition
\be
\label{eq:tensordtp}
\mathcal{Q}_{ij} = \langle \tr O_i \tr O_j \rangle = G_{i,j} \succeq 0 \, .
\ee 
Imposing the two conditions above \eqref{eq:tensorGramcond} and \eqref{eq:tensordtp}, as well as the Schwinger-Dyson equations \eqref{eq:tensor_sdes} and traceless component positivity ($\mathcal{M} - \mathcal{Q} \succeq 0$), we examined constraints on the present system \eqref{eq:actionpillow}. In this case, we found that the present tensor model \eqref{eq:actionpillow} has a key advantage over matrix models. Since the Schwinger-Dyson equations \eqref{eq:tensor_sdes} depend explicitly on the order $D$ of the tensor and its size $N$, these bounds vary as we change the values of $D$ and $N$.

\begin{figure}[h]
	\centering
	\begin{minipage}{0.32\textwidth}
		\centering
		\includegraphics[width=\linewidth]{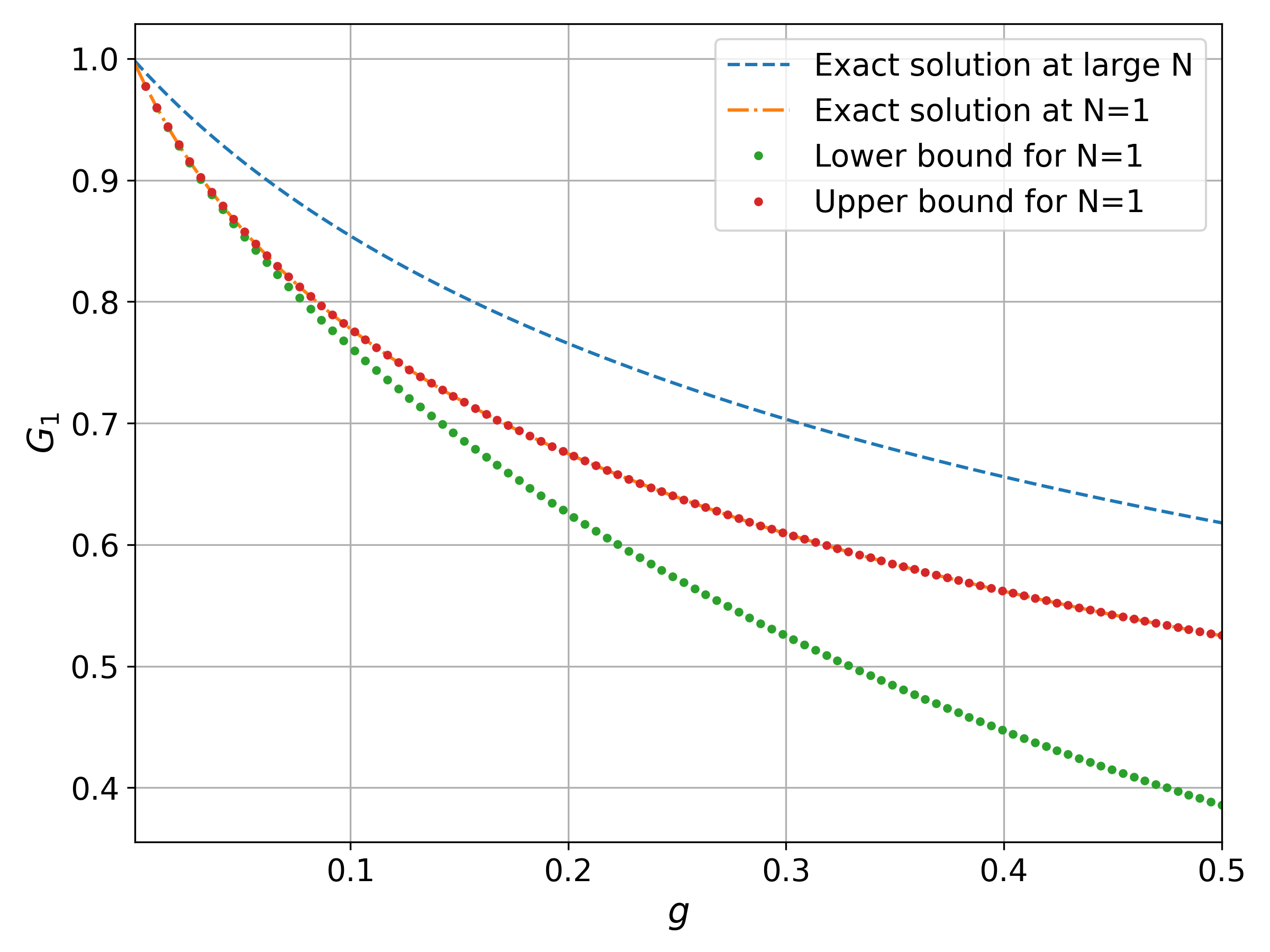}
	\end{minipage}\hfill
	\begin{minipage}{0.32\textwidth}
		\centering
		\includegraphics[width=\linewidth]{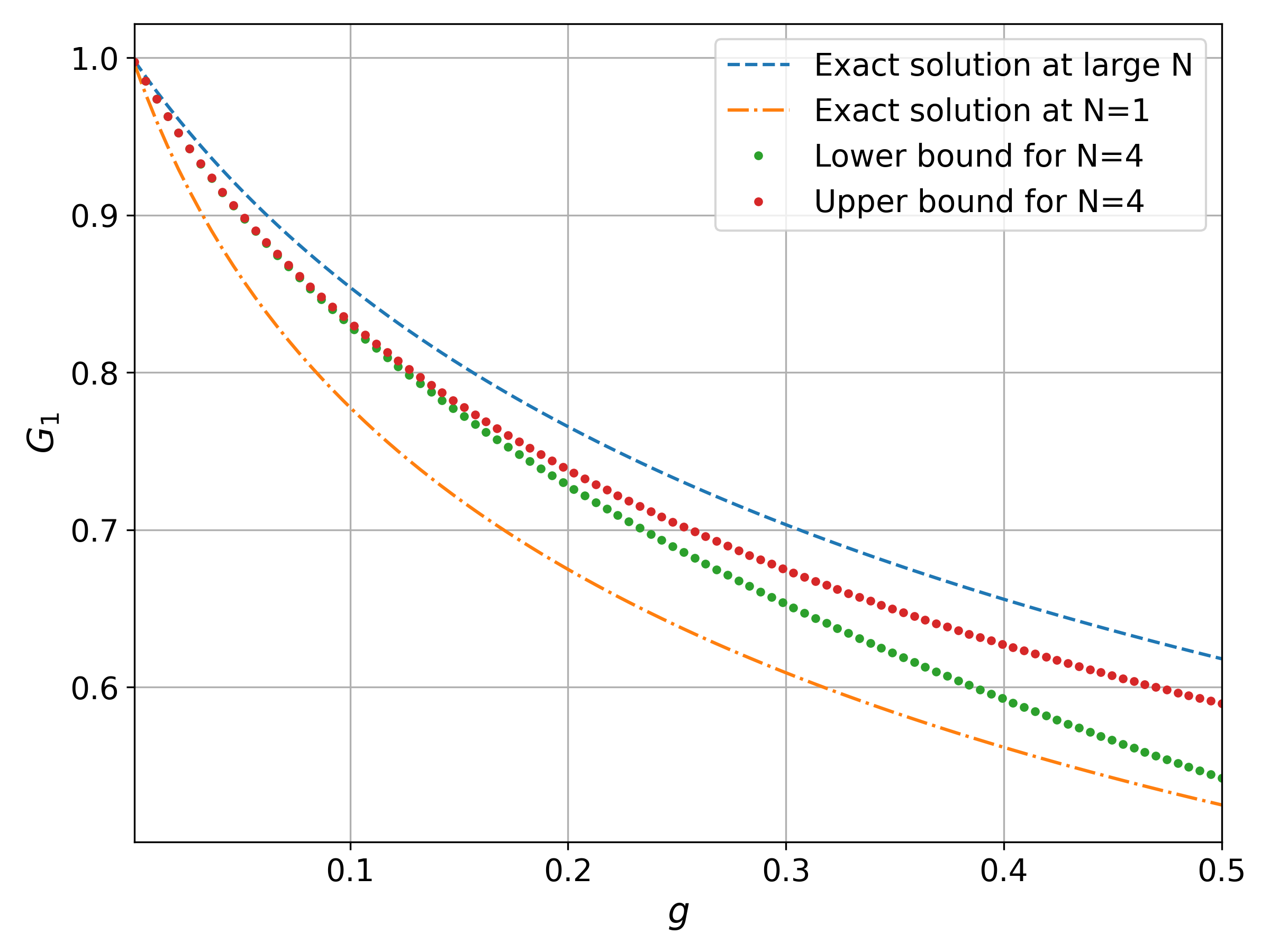}
	\end{minipage}\hfill
	\begin{minipage}{0.32\textwidth}
		\centering
		\includegraphics[width=\linewidth]{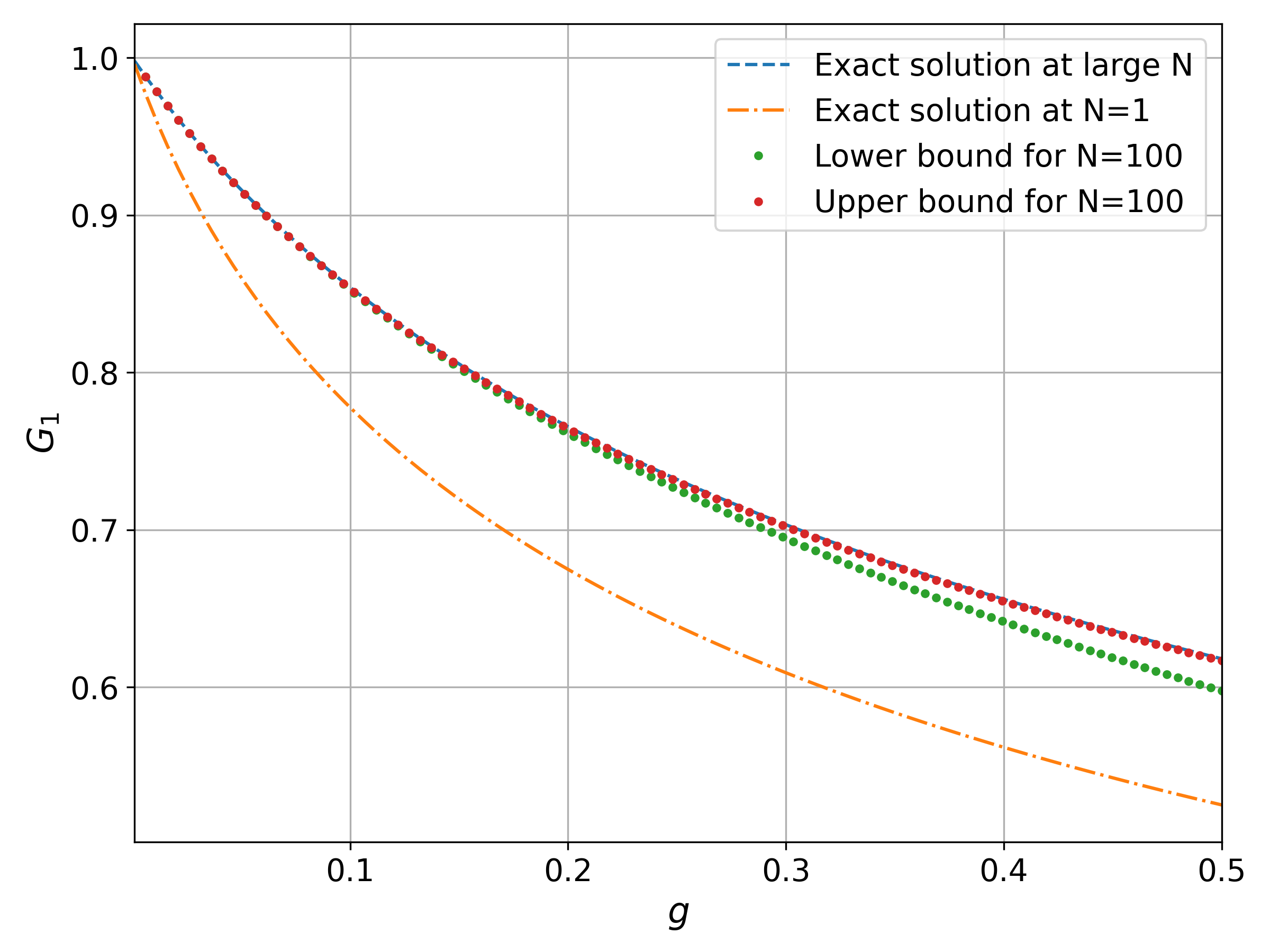}
	\end{minipage}
	\caption{Bounds for the two-point function of an order 3 tensor as a function of the coupling $g$ for $g \geq 0$. $N$ is chosen to be 1 (left), 4 (center), and 100 (right) respectively. The size of the Gram matrix is $4 \times 4$.
    }
	\label{fig:rank3_gpos}
\end{figure}

To illustrate how this dependence manifests itself, let us first study the case of an order $D=3$ tensor with positive values of the coupling $g$ (see Figure \ref{fig:rank3_gpos}). For this specific case, we also imposed the additional constraints $G_i \geq 0 ,, \forall , i$ in order to obtain stronger lower bounds. These constraints follow directly from the observation that $G_i \equiv \big\langle \tr (T T^\dagger)^i \big\rangle$ must be a positive quantity. Without these constraints, the bounds allow for negative values of $G_i$, which is unphysical. For $N = 1$, the upper bound of the constraint region is consistent with the exact solution of the model for $N = 1$. As we increase $N$, the region then gets uplifted in the intermediate values located between the exact solution for $N = 1$ and large $N$. The center plot in Figure \ref{fig:rank3_gpos} shows $N = 5$ as an example case. Finally, at large values of $N$, the upper bound of the constraint region becomes consistent with the exact solution at large $N$. The right plot in Figure \ref{fig:rank3_gpos} shows $N = 100$ as an example of this limit.

\begin{figure}[h]
	\centering
	\begin{minipage}{0.32\textwidth}
		\centering
		\includegraphics[width=\linewidth]{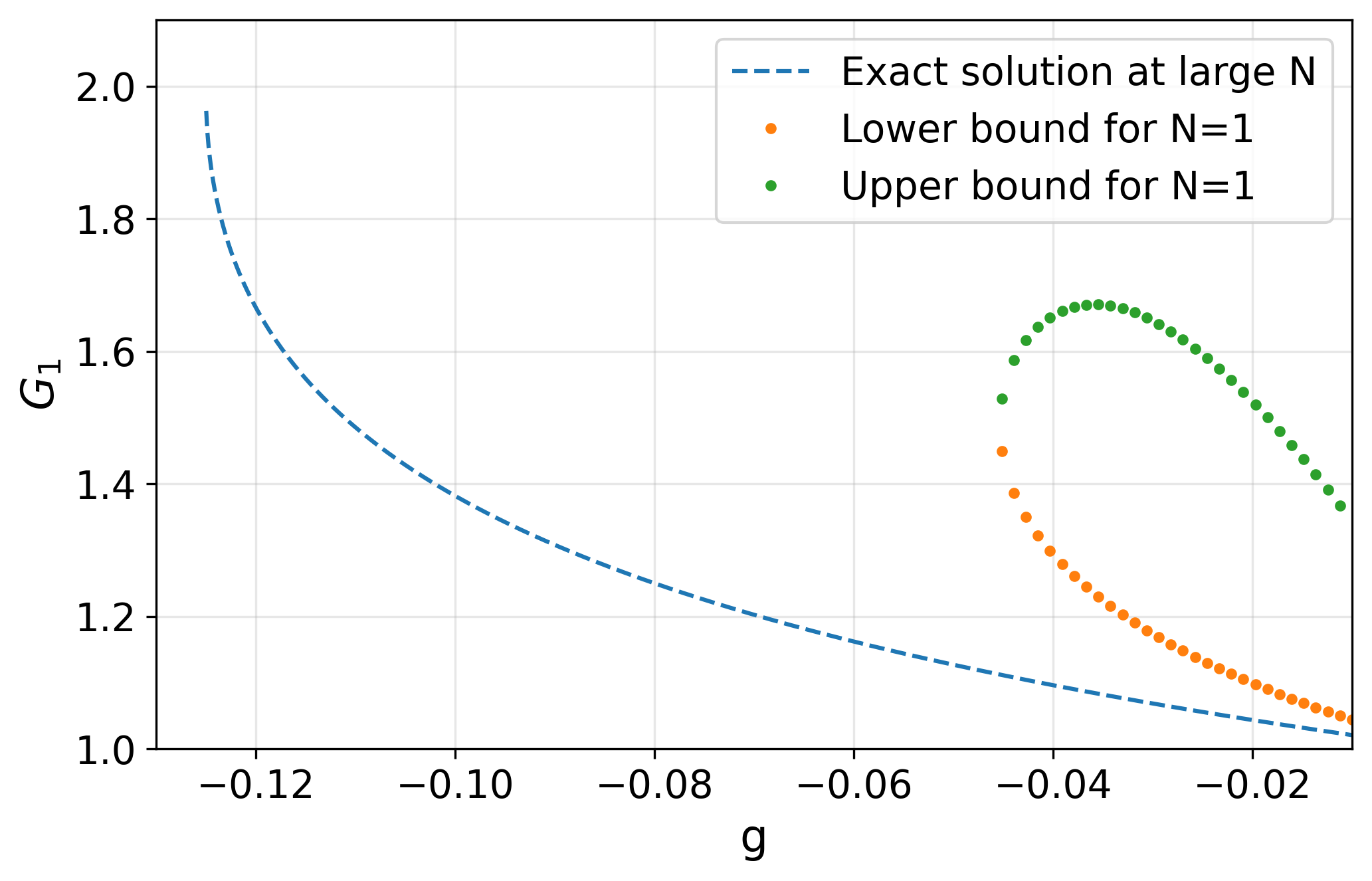}
	\end{minipage}\hfill
	\begin{minipage}{0.32\textwidth}
		\centering
		\includegraphics[width=\linewidth]{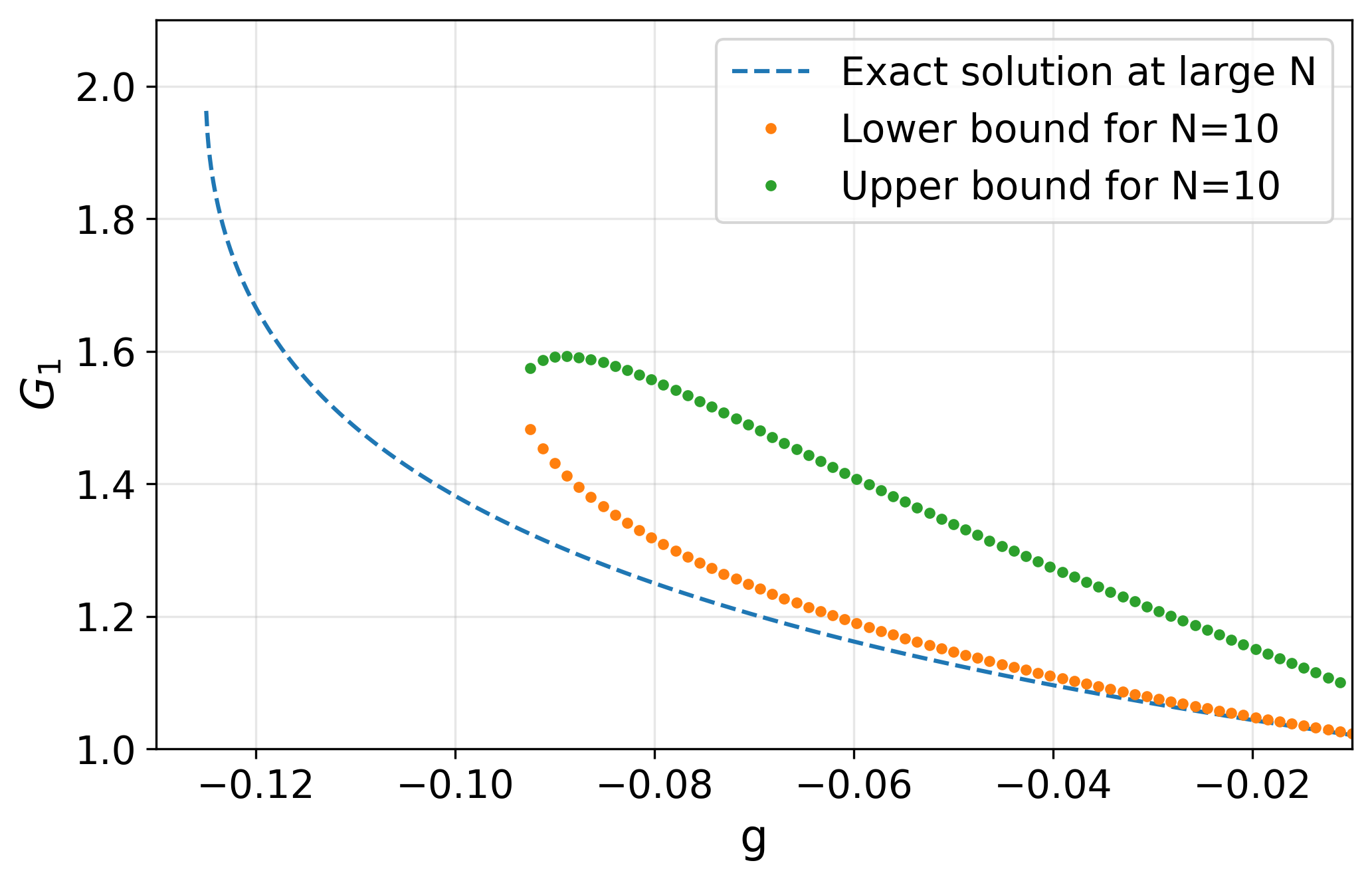}
	\end{minipage}\hfill
	\begin{minipage}{0.32\textwidth}
		\centering
		\includegraphics[width=\linewidth]{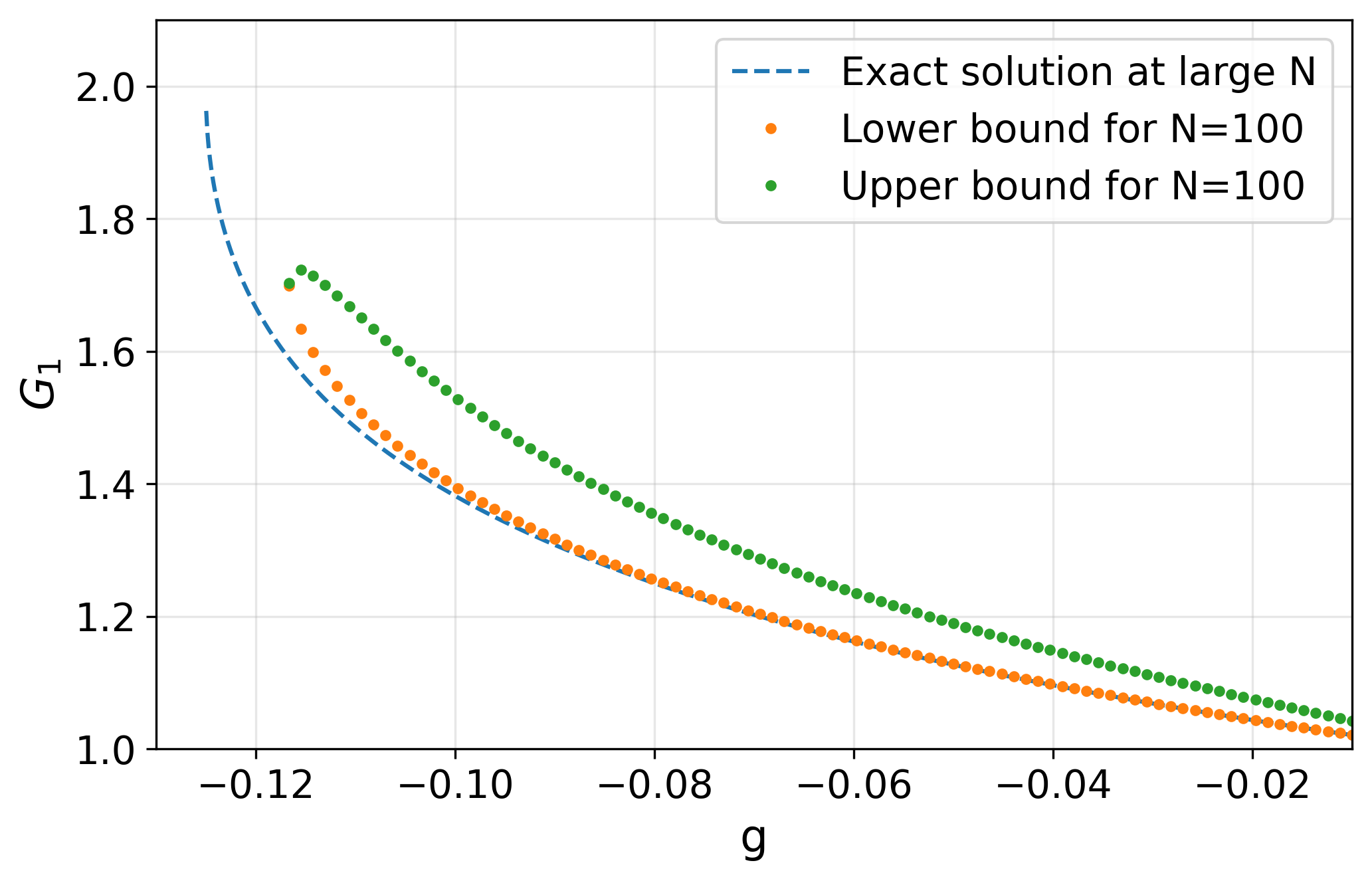}
	\end{minipage}
	\caption{Bounds for the two-point function of the order 3 tensor model as a function of negative values of the coupling $g$. $N$ is chosen to be 1 (left), 10 (center), and 100 (right) respectively. The size of the Gram matrix is $4 \times 4$.}
	\label{fig:rank3_gneg}
\end{figure}

For $D=3$, we also examined the behavior of the bounds when the coupling $g$ is negative (see Figure \ref{fig:rank3_gneg}). At finite $N$, the behavior of this model for $g < 0$ is unknown to the authors of the present paper. However, in the large $N$ limit, one expects the bounds to become consistent with \eqref{eq:G1_largeN}. This is what we indeed observe. For $N = 1$ (left plot in Figure \ref{fig:rank3_gneg}), we find bounds that are inconsistent with the exact solution in the large $N$ limit. However, as we increase $N$, the lower bound of the constraint region becomes increasingly consistent with the large $N$ solution. In Figure \ref{fig:rank3_gneg}, the cases $N = 10$ and $N = 100$ provide examples of how the bounds become increasingly consistent with the exact solution at large $N$.

\begin{figure}[h]
	\centering
	\begin{minipage}{0.48\textwidth}
		\centering
		\includegraphics[width=\linewidth]{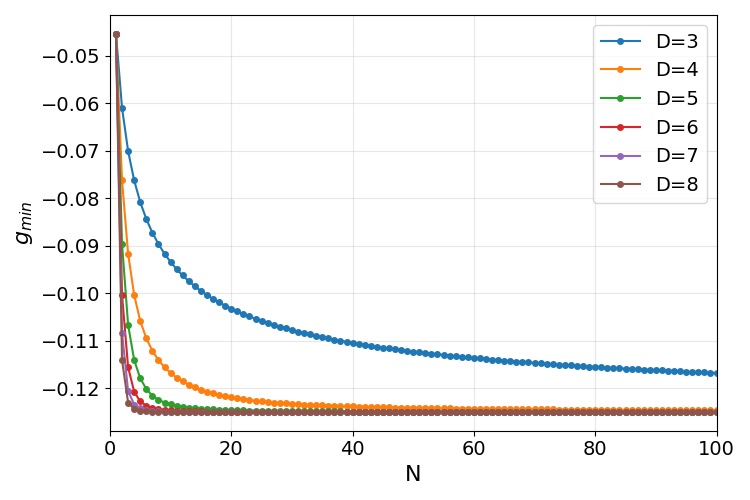}
	\end{minipage}\hfill
	\begin{minipage}{0.48\textwidth}
		\centering
		\includegraphics[width=\linewidth]{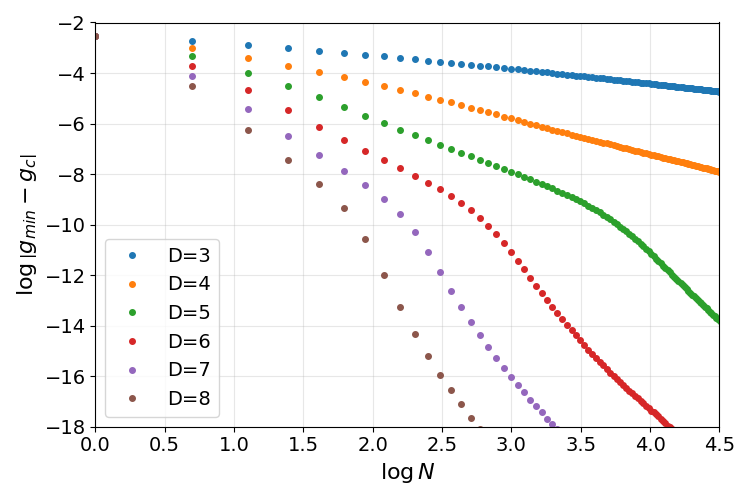}
	\end{minipage}
	\caption{(Left) Lower bound of the minimum allowed value of the coupling $g_{min}$ as a function of the dimension of the tensor $N$. (Right) Log-log plot of $|g_{min}-g_c|$ as a function of $N$. The size of the Gram matrix is $4 \times 4$.}
	\label{fig:gmin_vs_N}
\end{figure}

For higher order tensors, we obtain bounds similar to the order 3 case. The main difference is that the bounds reach the large $N$ limit much faster. We attribute this faster convergence to the $1/N^{D-2}$ scaling of some terms on the left-hand side of the Schwinger-Dyson equations for the present model  \eqref{eq:tensor_sdes}. As one increases $D$, the terms which scale as $1/N^{D-2}$ become subdominant faster as a function of $N$, making the bounds converge to the large $N$ limit faster. In Figure \ref{fig:gmin_vs_N}, we show this phenomenon by plotting the lower bound on the quartic coupling $g$ as a function of $N$ found using bisection refinement (see Appendix \ref{sec:bisec_ref} for more details). We see that, for $D \geq 3$, the minimum allowed value $g_{min}$ of the quartic coupling $g$ asymptotes to $g_c = -1/8 = -0.125$ in the large $N$ limit. Moreover, we find that this limit is achieved faster as one increases the value of $D$. For $D = 3$, values of $N$ larger than 100 are required in order to reach this limit. In comparison, for $D \geq 4$, values of $N$ less than 50 seem to be sufficient. 

In Figure \ref{fig:gmin_vs_N}, we also compared the behavior of $\log g_{min}$ as a function of $\log N$. In this plot, the curves for $D = 3, 4, 5, 6, 7$ and $8$ show interesting linear regimes and inflection points. It is tempting to speculate that these regimes might be related to different scaling limits of the theory. However, it is so far too soon to say if these linear regimes are related to physical processes or if they are numerical artifacts. Sharper bounds may be able to clarify this question in future work.

Like for the matrix model case, we found that the bounds did not improve in strength as the size of the $n \times n$ Gram matrix was taken beyond $n = 4$. This might be again because beyond $n = 4$, unconstrained values of $G_{n,m}$ prevent improvements of the constraints. For $n \leq 4$, all values of $G_{n,m}$ that appear in $\mathcal{M}$ also appear in $\mathcal{Q}$, and are therefore constrained by the inequalities $\mathcal{M} \succeq 0$, $\mathcal{Q} \succeq 0$, and  $\mathcal{M} - \mathcal{Q} \succeq 0$. However, for $n \geq 4$, some values of $G_{n,m}$ that appear in $\mathcal{M}$ do not appear in $\mathcal{Q}$. For this reason, they become only constrained by the inequalities $\mathcal{M} \succeq 0$ and $\mathcal{M} - \mathcal{Q} \succeq 0$, giving rise to weaker constraints. Like for the matrix case, we expect the bounds to improve with additional constraints.

\section{Conclusion}

We explored how matrix bootstrap techniques can be used to constrain matrix and tensor models at finite $N$ using a Gaussian model with a quartic interaction as example. For matrix models, the bounds we considered do not seem sufficient to obtain bounds at arbitrary $N$. With or without properly weighting single-trace quantities by a factor of $1/N$, we found that bounds did not change as a function of $N$. Rather, they seem to allow for all possible curves located between the $N = 1$ and the large $N$ limit. It seems, however, that it is possible to obtain bounds related to the $N = 1$ and large $N$ limit. To do so, one must impose double-trace factorization properties that apply to the $N = 1$ or large $N$ limit. For example, imposing $m_{k,l} = m_k m_l$, which follows from large $N$ factorization, one obtains bounds only consistent with the large $N$ limit. Conversely, if we impose $m_{k,l} = m_{k+l}$, which is valid for $N = 1$ from the $x^k x^l = x^{k+l}$, we obtain bounds only consistent with the $N = 1$ limit. For tensor models, we found bounds at arbitrary $N$ without having to rely on properties of double-trace expectation values. For $N = 1$, these bounds are consistent with the exact solution in the $N = 1$ limit. Increasing $N$, one then finds intermediate bounds located between the $N = 1$ and large $N$ limit that become progressively consistent with the exact large $N$ limit as one increases the value of $N$.

Our results bring further evidence that properties of multi-trace expectation values play a key role in obtaining bounds for matrix models at finite $N$. This is consistent with earlier work \cite{Kazakov:2024ool}, in which it was found that multi-trace properties allow bounds for SU(2) and SU(3) models. Such trace properties may help strengthen present bounds at finite $N$.

The present results raise interesting questions related to the study of matrix and tensor models. Notably, the tensor model we studied in the present paper had some notable matrix-like properties that greatly simplified the problem. It would be interesting to study broader classes of tensor models to see if interesting bounds can be extracted. The next simplest model to study would be one described by the tensor potential \eqref{eq:actionpillow}, but where the complex tensor $T = T_{A B}$ has two fat indices $A = (a_1 , a_2 , ... a_n)$ and $B = (a_{n+1} , a_{n+2} , ... a_D)$. In this case, the Schwinger-Dyson equation would probably be very similar to \eqref{eq:tensor_sdes}. However, we expect the bounds to evolve differently as a function of $N$. 

Another interesting avenue of research could be to explore if tensor models admit a continuum limit in the finite $N$ regime using bootstrap methods. To see how this continuum regime could come about, let us take one of the simplest forms of a tensor model is taken, e.g., presented in \cite{Bonzom:2011zz} for example, where one assigns each $1$-cell the lattice spacing $a$. In this case, we can match the tensor model action to the discretized Einstein-Hilbert action, and identify the relation between Newton's constant $G$ and the dimension of the tensor $N$ to be \cite{Gurau:2016cjo}
\be
\frac{G}{a^{D-2}} = {\text {constant}} 
\, 
\frac{1}{\ln N}
\,.
\ee
From this identification, in order to address quantum gravity from tensor model perspective, one may be interested in not necessarily restricting ourselves to the limit $N \rightarrow \infty$, but rather may be interested in tuning to some finite $N$ so as to reach the experimentally measured value of $G$. Of course, tuning to criticality will still be necessary to achieve such limit. We leave this avenue of research for future work.

\section*{Acknowledgements}

The authors would like to acknowledge Dario Benedetti, Zechuan Zheng, and Vasily Sazonov for meaningful discussions.

\appendix

%\addcontentsline{toc}{section}{Appendix}
\addtocontents{toc}{\protect\contentsline{section}{Appendix}{}{}}
\section*{Appendix}

\section{Coupling re-scaling trick}
\label{sec:coupling_rescaling}

When finding bounds using semi-definite programming methods, numerical instabilities can occur when correlator coefficients differ from each other by a large factor. For the models studied in the present paper, these instabilities tend to appear in the small $g$ regime, where some entries of the matrices $\mathcal{M}$ and $\mathcal{Q}$ become disproportionately large compared to one another. To see how this is the case, let us consider the matrix model with a quartic interaction considered in Section \ref{sec:matrix_boot}. In this case, a $4 \times 4$  Gram matrix takes the form
\be
\mathcal{M} = 
\begin{pmatrix}
	1 & 0 & m_2 & 0\\
	0 & m_2 & 0 & \frac{1 - m_2}{g} \\
	m_2 & 0 & \frac{1 - m_2}{g} & 0 \\
	0 & \frac{1 - m_2}{g} & 0 & \frac{(-1 + (1 + 2 g) \, m_2 + g \, m_{1,1})}{g^2} 
\end{pmatrix}
\, .
\ee
when recursively substituting $m_k$ for $k > 2$ for expressions involving lower power expectation values using the Schwinger-Dyson equations. In the lower right portion of the Gram matrix, the elements of the matrix have factors of $1/g$ and $1/g^2$, which blow up as $g \rightarrow 0$ and introduce numerical instabilities. To help with this issue, one can re-scale the Gram matrix $\mathcal{M}$ using the transformation
\be
\mathcal{M} \rightarrow \mathcal{D} \mathcal{M} \mathcal{D}
\ee
where $\mathcal{D}$ is a diagonal matrix of the form
\be
\mathcal{D} = \text{diag}(1, g , g, ..., g ) \, .
\ee
This re-scaling introduces extra factors of $g$ in the lower right section of the matrix, hence taming the $1/g$ and $1/g^2$ divergences. From the point of view of imposing that the matrix $\mathcal{M}$ is positive semi-definite
\be
v^T \mathcal{M} v \geq 0 \quad \forall v \, ,
\ee
the transformation $\mathcal{M} \rightarrow \mathcal{D} \mathcal{M} \mathcal{D}$ preserves semi-definiteness as it is equivalent to the change of variables $v \rightarrow \mathcal{D} v$. For this reason, one can proceed with the present transformation without changing the results of the numerical optimization problem, and improving convergence while doing so.

\section{Bisection Refinement}
\label{sec:bisec_ref}

To find the value lower bounds on the coupling $g$ as a function of the tensor size $N$, we used the method of bisection refinement to determine the lowest value of $g$ for which SDPA converges. This method functions as follows. First, we determine an interval $[g_1 , g_{min}]$ where we expect SDPA to stop converging for some value $g_{min}$ such that $g_1 < g_{min} < g_2$, and converge in the interval $[g_{min} , g_2]$. We then test if the algorithm converges for $g = \frac{g_2-g_1}{2}$. If it converges, this means that $g_{min}$ lies in the interval $[g_1 , g]$. In this case, we repeat the process in the interval $[g_1 , g]$. If it does not converge, this means that $g_c$ lies in the interval $[g , g_2]$. In this case, we repeat the process in the interval $[g , g_2]$. The process is repeated $n$ times until the desired accuracy is achieved. For the present analysis, we found $n = 20$ to be sufficient to obtain the results in Figure \ref{fig:gmin_vs_N}.

\bibliography{cite.bib}
\bibliographystyle{JHEP.bst}

\end{document}